\documentclass[%
 aip,
 amsmath,amssymb,
 reprint,%
]{revtex4-1}

\usepackage{graphicx}
\usepackage{dcolumn}
\usepackage{bm}

\usepackage[utf8]{inputenc}
\usepackage[T1]{fontenc}
\usepackage{mathptmx}
\usepackage{etoolbox}
\usepackage{physics}
\usepackage{booktabs}
\usepackage{xcolor}
\usepackage{algorithm}
\usepackage{algpseudocode}
\usepackage{tablefootnote}
\usepackage{multirow}
\usepackage{hyperref}
\hypersetup{
    colorlinks=true,
    linkcolor=blue,
    filecolor=magenta,      
    urlcolor=cyan,
}

\definecolor{tnteal}{rgb}{0,50,28}

\makeatletter
\def\@email#1#2{%
 \endgroup
 \patchcmd{\titleblock@produce}
  {\frontmatter@RRAPformat}
  {\frontmatter@RRAPformat{\produce@RRAP{*#1\href{mailto:#2}{#2}}}\frontmatter@RRAPformat}
  {}{}
}%
\makeatother
\begin{document}

\preprint{AIP/123-QED}

\title{A novel coupled-cluster singles and doubles implementation that combines the exploitation of point-group symmetry and Cholesky decomposition of the two-electron integrals}

\author{Tommaso Nottoli}%
\affiliation{Dipartimento di Chimica e Chimica Industriale, Universit\`{a} di Pisa, Via G. Moruzzi 13, I-56124 Pisa, Italy}

\author{J{\"u}rgen Gauss}%
 \email{gauss@uni-mainz.de}
\affiliation{Department Chemie, Johannes Gutenberg-Universit{\"a}t Mainz, Duesbergweg 10-14, D-55128 Mainz, Germany}

\author{Filippo Lipparini}%
 \email{filippo.lipparini@unipi.it}
\affiliation{Dipartimento di Chimica e Chimica Industriale, Universit\`{a} di Pisa, Via G. Moruzzi 13, I-56124 Pisa, Italy}

\date{\today}

\begin{abstract}
A novel implementation of the coupled-cluster  singles and doubles (CCSD) approach is presented that is specifically tailored for the treatment of large, symmetric systems. It fully exploits Abelian point-group symmetry and the use of the Cholesky decomposition of the two-electron repulsion integrals. In accordance with modern CCSD algorithms, we propose two alternative strategies for the computation of the so-called particle-particle ladder term. The code is driven towards the optimal choice depending on the available hardware resources. As a large-scale application, we computed the frozen-core correlation energy of buckminsterfullerene (C$_{60}$) with a polarized valence triple-zeta basis set (240 correlated electrons in 1740 orbitals). 
\end{abstract}

\maketitle

The accurate and reliable determination of the energy and properties of a molecular system is a pivotal aspect of quantum chemistry. In this context, coupled-cluster (CC) methods\cite{vcivzek1966correlation,vcivzek1969correlation,Shavitt09}
are an optimal choice due to their remarkable accuracy and size-extensivity. For many applications, CC with singles and doubles excitations (CCSD)\cite{Purvis82} offers a good compromise between accuracy and computational cost. Nevertheless, its routine use is restricted to small- to medium-sized systems. The computational cost scales in fact with the sixth power of the number of basis functions, and the storage requirements are high due to the fact that two-electron integrals involving three and four virtual orbitals need in principle to be stored. 

To overcome such computational limitations various strategies have been developed and are often combined together to get the most efficient implementation. A first class of strategies aims at reducing the overall scaling of the calculation by enforcing the local nature of electron correlation.\cite{hampel1996local,schuetz2001,scuseria1999linear,subotnik2006,auer2006,neese2009efficient,neese2009efficient2,ma2018explicitly} Other low-scaling solutions are given by methods that compress the double-excitation amplitudes\cite{kinoshita2003singular} as for example in the tensor hypercontraction (THC) scheme,\cite{parrish2012tensor,parrish2013exact,parrish2019rank} or by real-space methods.\cite{Bischoff17a} Using such techniques, impressive large-scale calculations can be done.\cite{riplinger2013natural} However, reduced-scaling methods come with various difficulties, that include the use of cutoffs to enforce the locality of correlation, complicated expressions for analytical derivatives,\cite{Gauss00,Rauhut01,datta2016analytic,pinski2019analytical,stoychev2021dlpno} 
and overall accuracy control that requires some user experience. 
A different family of strategies does not reduce the formal scaling of the method, but allows for very efficient implementations that can be vectorized, parallelized, and formulated mostly in terms of highly optimized matrix-matrix multiplications. This is achieved by introducing a low-rank approximation of the two-electron repulsion integrals (ERI), such as density fitting (DF)\cite{Whitten1973,Dunlap79,Vahtras93,Feyereisen1993,Eichkorn95} and Cholesky decomposition (CD). \cite{Beebe1977,Roeggen1986,Koch2003,Roeggen2008,Aquilante2011,Epifanovsky2013,Folkestad2019,zhang2021toward,pedersen2023versatility} 
Such techniques can be used to reduce the prefactor in the overall method complexity, which remains $\mathcal{O}(N^6)$, and can drastically lower integral storage and bandwith requirements and improve the parallelization potential.

Simultaneously and in parallel with this, new CC implementations also focus on deploying efficient computational realizations that leverage the full potential of modern computer architectures and utilize parallel computing. Recently, various works that combine massively parallel implementation strategies -- based on the joint effort given by the message-passing interface (MPI) and OpenMP -- with or without DF/CD, showed impressive calculations on large systems without exploiting local correlation.\cite{harding08,prochnow10,pitovnak2011parallelized,asadchev2013fast,deumens2011software,kaliman2017new,janowski2008efficient,peng2016massively,shen2019massive,solomonik2014massively,gyevi2019integral,datta2021massively}  To the best of our knowledge the largest CC 
calculation, actually at the CCSD(T)\cite{raghavachari1989fifth} level, was done with as many as 1624 orbitals (without the use of point-group symmetry) using 288 cores.\cite{datta2021massively} 

In this work, we present an implementation of CCSD that exploits the CD of the ERI matrix. The main element of novelty in our implementation is that it fully exploits Abelian point-group symmetry. The implementation is highly vectorized, as it mostly relies on efficient level 3 BLAS routines to perform the bulk of the floating-point operations, and is parallelized using OpenMP. It is therefore well suited to treat, on accessible computational resources, medium-sized molecular systems and in particular large symmetric molecules. We choose CD instead of DF because, though DF affords a more compact representation of the ERIs, its accuracy depends on the fitted auxiliary basis set that is used for the calculation and cannot be determined \emph{a priori} nor improved in a systematic way. On the contrary, the accuracy of the CD representation is rigorously controlled by the threshold used to truncate the decomposition, which is the only user-defined parameter that controls the calculation. Our implementation has been realized in a development version of the \textsc{CFour} suite of programs.\cite{cfour,matthews2020coupled}

In the following we adopt the usual convention for the indices: $\mu, \nu, \sigma, \dots $ refer to atomic orbitals; $p, q, r, \dots$ refer to generic molecular orbitals; $i, j, k, \dots$ refer to occupied orbitals; $a, b, c, \dots $ refer to virtual orbitals; and $P, Q, \dots $ refer to the vectors of the Cholesky basis. To get the CCSD correlation energy for closed-shell systems
\begin{equation}
    \Delta E = 2 \sum_{ia}t_i^af_{ai} + \sum_{ij}\sum_{ab}\Tilde{\tau}_{ij}^{ab}\bra{ij}\ket{ab},
\end{equation}
a set of projected non-linear equations needs to be solved
\begin{align}
    &\bra{\Phi_i^a}\Bar{H}\ket{0} = 0, \\
    &\bra{\Phi_{ij}^{ab}}\Bar{H}\ket{0} = 0,
\end{align}
where $f_{ai}$ is the $ai$ element of the Fock matrix; $\Tilde{\tau}_{ij}^{ab} = 2\tau_{ij}^{ab}-\tau_{ji}^{ab}$ is the spin-adapted tau amplitude, which is the usual combination of double- ($t_{ij}^{ab}$) and single-excitation ($t_i^a$) amplitudes and in the closed-shell case assumes the following form $\tau_{ij}^{ab} = t_{ij}^{ab} + t_i^at_j^b$; $\bra{ij}\ket{ab}$ is a two-electron integral, written using the Dirac convention; $\ket{0}$ is the reference Hartree-Fock (HF) determinant; $\ket{\Phi_i^a}$ and $\ket{\Phi_{ij}^{ab}}$ are singly- and doubly-excited determinants, respectively; and $\Bar{{H}} = e^{-T}He^T$ is the similarity-transformed Hamiltonian. The algebraic expressions can be found elsewhere. In particular, we have implemented the spin-restricted closed-shell equations\cite{scuseria1988efficient}. We formulated the equations using spin-adapted forms of the $\mathcal{F}$ and $\mathcal{W}$ intermediates first introduced in the spin-orbital basis by Stanton, Gauss, and coworkers.\cite{stanton1991direct,gauss1991coupled}

The CD of the ERIs (in Mulliken notation) reads:
\begin{equation}
    \bra{\mu\nu}\ket{\rho\sigma} = \sum_P L^P_{\mu\rho}L^P_{\nu\sigma},
\end{equation}
where $L^P_{\mu\rho}$ denotes an element of the $P$th Cholesky vector (CV). In order to account for Abelian point-group symmetry the CD is done using symmetry-adapted linear combinations of atomic orbitals. The resulting CVs are classified by means of the irreducible representation to which the index $P$ belongs ($\Gamma_P$). For the totally-symmetric vectors we store only the lower triangular block while for the others we store only the sub-blocks for which $\Gamma_{\mu} \leq \Gamma_{\rho} = \Gamma_P \otimes \Gamma_\mu$. The CVs are transformed to the molecular orbital basis at the beginning of the CC calculation. Furthermore, in our implementation we always assemble and keep in core the integrals with at most two virtual indices such as $\bra{ab}\ket{ij}$ and $\bra{ij}\ket{kl}$. 

The exploitation of symmetry enables a reduction of the memory requirement and the number of floating-point operations by approximately a factor $h^2$, $h$ being the order of the point-group. Given a generic contraction between two four-index arrays, it is performed using the direct-product decomposition (DPD) approach\cite{stanton1991direct} which means that the contraction is subdivided into $h$ distinct operations. The rate-determining step in a CCSD code is given by the so-called particle-particle ladder (PPL) term
\begin{equation}
    Z_{ij}^{ab} = \sum_{ef}\tau_{ij}^{ef}\mathcal{W}_{abef}.
\end{equation}
The $\mathcal{W}_{abef}$ intermediate can be written in terms of CVs as
\begin{equation}\label{eq:w_abef}
    \mathcal{W}_{abef} = \sum_P \left[\left(L^P_{ae}-t^P_{ae}\right)L^P_{bf} - \sum_m t_m^b L^P_{ae}L^P_{mf}\right],
\end{equation}
and $t^P_{ab} = \sum_m t_m^aL^P_{bm}$. To reduce the computational cost from its formal $\frac{1}{2}O^2V^4$ scaling to $\frac{1}{4}O^2V^4$ we exploited the symmetric/antisymmetric algorithm,\cite{saebo1987fourth,scuseria1988efficient} which was also successfully exploited in other recent CD/DF-CC implementations.\cite{deprince2013accuracy,bozkaya2016analytic,schnack2022efficient} $O$ and $V$ denote the number of occupied and virtual orbitals, respectively. In particular, $Z_{ij}^{ab}$ is computed as the sum of a totally symmetric ($S$) and an antisymmetric ($A$) contribution:
\begin{equation}
    Z_{ij}^{ab} = S_{ij}^{ab} + A_{ij}^{ab},
\end{equation}
where
\begin{align}
    & S_{ij}^{ab} = \sum_{ef} {}^+\tau_{ij}^{ef}{}^+\mathcal{W}_{abef}, \\
    & A_{ij}^{ab} = \sum_{ef} {}^-\tau_{ij}^{ef}{}^-\mathcal{W}_{abef},
\end{align}
and the $+$ and $-$ denote symmetric and antisymmetric combinations, respectively.

Moreover, we note that the $\mathcal{W}_{abef}$ intermediate can be rewritten as a single contraction
\begin{equation}\label{eq:w_abef_t1}
    \mathcal{W}_{abef} = \sum_P \left(L^P_{ae}-t^P_{ae}\right)\left(L^P_{bf}-t^P_{bf}\right).
\end{equation}
However, the following contribution 
\begin{equation}\label{eq:new-hhl}
    Y_{ij}^{ab} = \sum_{ef}\sum_{mn} \tau_{ij}^{ef}t_m^at_n^b\left(\sum_P L^P_{me}L^P_{nf}\right)
\end{equation}
needs to be subtracted from the double-excitation equations, as it is spuriously included in Eq.~\eqref{eq:w_abef_t1}. This is in practice not a problem, as Eq.~\eqref{eq:new-hhl} introduces a hole-hole ladder-like term, which only requires $O^4V^2$ operations and $O^2V^2$ words of memory for its computation. The price of computing such a term is compensated by not having to take into account single-excitation amplitudes in the evaluation of the PPL term, as in Eq.~\eqref{eq:w_abef}.
This approach can thus be seen as a simple tool to slightly reduce the operation count and memory demand of the PPL contraction subroutine when the T$_1$-transformed Hamiltonian 
is not used.\cite{koch1994t1} We note that the quantity $L^P_{ae}-t^P_{ae}$ is not symmetric with respect to the interchange of indices $a$ and $e$.

An efficient implementation of the PPL term should take into account memory issues, and it should distribute the right amount of data to the CPU in order to retrieve its peak performance. These two considerations require in general the adoption of different strategies that are often in conflict with respect to each other. As the main cornerstone, we avoid to assemble and thus to store $V^4$ and $V^3O$ arrays; for this reason only batches of integrals are assembled and contracted on-the-fly. In particular, we implemented two different strategies for the PPL term. In the first version, which is outlined in Algorithm \ref{loop_a}, we fix index $a$ and distribute the operations over shared-memory threads using the OpenMP paradigm; therefore, the memory required to store temporary arrays is $V^3N_{threads}$. Here we used for $\mathcal{W}$ the expression reported in Eq.~\eqref{eq:w_abef}. This algorithm has the advantage of computing the PPL contraction with few, large matrix-matrix multiplications that are efficiently performed using the level 3 BLAS DGEMM routine, but as the number of threads increases, it can become too demanding in terms of memory requirements. We therefore implemented as a second option a different algorithm, which is shown in Algorithm \ref{loop_ab}, where we fix both the $a$ and $b$ indices and again distribute the operations over OpenMP threads. In this situation the memory required to store the intermediates scales as $V^2N_{threads}$. This time, Eq.~\eqref{eq:w_abef_t1} was used.

\begin{algorithm}[H]
\caption{\textsc{PPL: Loop over a}}\label{loop_a}
\begin{algorithmic}[1]
        \State Form ${}^\pm\tau_{ij}^{ab}, \quad\forall\quad{\rm irr}$ 
        \For{irr2 = 1, n\_irrep}\Comment{OpenMP parallelized}
        \For{a = 1, n\_vir(irr2)}\Comment{OpenMP parallelized}
            \State $I^a_{ebf} = \sum_P(L^P_{ae}-t^P_{ae})L^P_{bf}$ 
            \State $J^a_{feb} = I^a_{ebf} - \sum_m t_m^b\left(\sum_PL^P_{mf}L^P_{ae}\right)$
            \State ${}^{\pm}W^a_{feb} = \frac{1}{2}(J^a_{feb} \pm J^a_{efb})$
            \State Form $S_{ij}^{ab}$ and $A_{ij}^{ab}$
            \State Distribute in $Z_{ij}^{ab}$
        \EndFor
        \EndFor
\end{algorithmic}
\end{algorithm}

\begin{algorithm}[H]
    \caption{\textsc{PPL: Loop over a and b}}\label{loop_ab}
    \begin{algorithmic}[1]
        \State Form ${}^\pm\tau_{ij}^{ab}, \quad\forall\quad{\rm irr}$ 
        \For{irr2 = 1, n\_irrep}\Comment{OpenMP parallelized}
        \For{a = 1, n\_vir(irr2)}\Comment{OpenMP parallelized}
        \For{irr = 1, n\_nirrep}\Comment{OpenMP parallelized}
            \State irr1 $=$ irr $\times$ irr2
            \State \textbf{if} (irr1 < irr2) cycle
            \State b\_end = n\_vir(irr1)
            \State \textbf{if} (irr == 1) b\_end = a
            \For{b = 1, b\_end}\Comment{OpenMP parallelized}
                \State $I^{ab}_{ef} = \sum_P(L^P_{ae}-t^P_{ae})(L^P_{bf}-t^P_{bf})$
                \State ${}^{\pm}W^{ab}_{ef} = \frac{1}{2}(I^{ab}_{ef} \pm I^{ab}_{fe})$
                \State Form $S_{ij}^{ab}$ and $A_{ij}^{ab}$
                \State Distribute in $Z_{ij}^{ab}$
            \EndFor
        \EndFor
        \EndFor
        \EndFor
    \end{algorithmic}
\end{algorithm}

\begin{figure}
    \centering
    \includegraphics[width=0.45\textwidth]{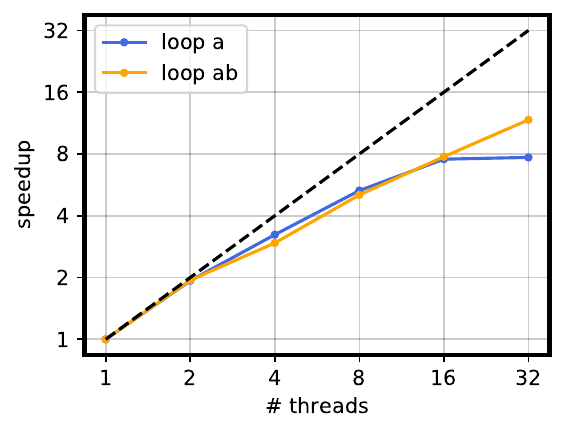}
    \caption{Speedup plot for CD  based CCSD/\textit{cc}-pVTZ calculations on coronene. For both algorithms we plot the ratio between iteration time for the serial calculation and the iteration time of the parallel ones. Both axes are in the $\log_2$ scale.}
    \label{fig:speedup}
\end{figure}

\noindent
To test the two PPL algorithms, we ran calculations on coronene (C$_{24}$H$_{12}$) enforcing $D_{\rm 2h}$ symmetry using Dunning's \textit{cc}-pVTZ basis set\cite{Dunning1989} and the frozen-core (fc) approximation, that is 888 orbitals of which 24 are kept frozen. The geometry is given as supplementary material in the Zenodo repository.\cite{nottoli_2023_ccsd_data} We note in passing that for non-symmetric systems our code performs similarly to the recent implementation presented by Schnack-Petersen {\sl et al.};\cite{schnack2022efficient} here, we focus on the performance of the implementation for symmetric molecules. All calculations were run on a single Intel Xeon Gold 6140M node equipped with 1.1 TB of RAM. The results are summarized in Table \ref{tab:coronene_results}. For both algorithms we run the calculation using 32 OpenMP threads, as expected the peak memory required is considerably reduced for the ``loop ab'' algorithm, and in particular it coincides with the memory required by the PPL subroutine. However, the timings for a single iteration are approximately a factor of two slower with respect to the implementation using the first algorithm. 
\begin{table}[]
    \centering
    \begin{tabular}{lcc}
         \toprule
         Algorithm & Memory (GB) & Time it. (min) \\
         \midrule
         loop a  & 272 & 3.5 \\
         loop ab & 47 & 6 \\
         \bottomrule
    \end{tabular}
    \caption{Summary of the comparison between the two implementations for the PPL contraction. The calculations are done on coronene with the \textit{cc}-pVTZ basis set using up to 32 OpenMP threads. We report for each algorithm type the peak-memory consumption in GB which coincides with the memory requested by the PPL subroutine, and the average time for a single CCSD iteration in minutes.}
    \label{tab:coronene_results}
\end{table}
We observe that both algorithms present an analogous speedup plot, as shown in Figure \ref{fig:speedup}, with the ``loop a'' algorithm performing slightly better for smaller numbers of threads but hitting a plateau sooner than the ``loop ab'' algorithm. 
This behavior is expected, as both algorithm execute in parallel a large number of \textsc{DGEMM} matrix-matrix multiplications, which make the code memory-bound. This in turns causes cache misses and other cache-related issue, making the cores idle while they are starving for more data to process. The ``loop ab'' algorithm scales slightly better because the matrix-matrix multiplications performed within the parallel loop are smaller, which reduce the memory bottleneck. Nevertheless, a somewhat satisfactory speedup factor of 8-10 can be achieved. Different parallelization strategies, as well as the deployment of the code on different architectures, will be investigated in the future.

\begin{figure}
    \centering
    \includegraphics[width=0.45\textwidth]{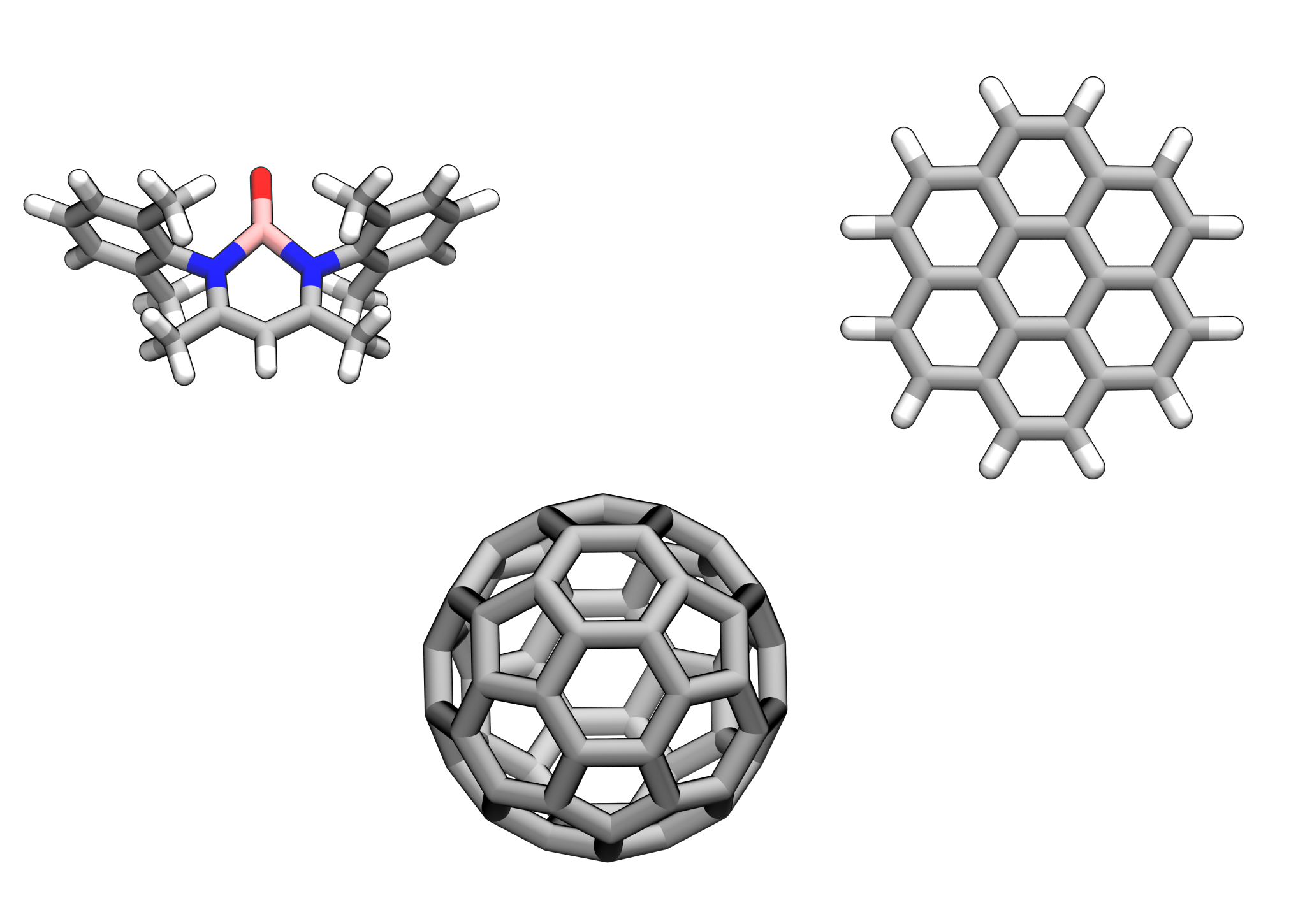}
    \caption{Molecular systems used to test the CD-CCSD code. From the left: (nacnac)Co$^{\rm III}$O, buckminsterfullerene, and coronene. Carbon is represented in silver, oxygen in red, nitrogen in blue, cobalt in pink, and hydrogen in white.}
    \label{fig:mol}
\end{figure}

The benefit obtained by exploiting point-group symmetry can be discerned by comparing the theoretical factor of reduction due to symmetry (FRS) with respect to the achieved FRS. Theoretical FRSs can be computed as the ratio between the total number of floating-point operations required for a specific contraction term in the CC equations with symmetry and the one without symmetry. Achieved FRSs are given by the ratio between the elapsed CPU time required for a specific contraction performed with and without enforcing Abelian symmetry. The results are summarized in Table~\ref{tab:FRS}. As test systems and basis sets we used corenene with (fc){\it cc}-pVTZ, buckminsterfullerene (C$_ {60}$) with (fc){\it cc}-pVDZ, and a cobalt $\beta$-diketiminato oxo complex [(nacnac)Co$^{\rm III}$O] with {\it cc}-pVDZ whose geometry was taken from Ref.~\citenum{conradie2007cobalt}. The three molecular representations can be seen in Figure~\ref{fig:mol}.
\begin{table}[h]
    \centering
    \begin{tabular}{ccccc} 
    \toprule
         &  \multicolumn{2}{c}{theoretical FRS}&  \multicolumn{2}{c}{achieved FRS}\\ 
         \cmidrule{2-3}\cmidrule{4-5}
         Molecule&  $O^3V^3$&  $O^2V^4$&  $O^3V^3$&    $O^2V^4$\\ 
         \midrule
         C$_{24}$H$_{12}$& 59 & 59 & 35~(47) & 10~(23)\\
         C$_{60}$& 64 & 63 & 44~(57) & 15~(28) \\
         (nacnac)Co$^{\rm III}$O& 16 & 15 & 12~(14) & 6~(9)\\ 
         \bottomrule
    \end{tabular}
    \caption{We report the theoretical and achieved factor of reduction due to symmetry (FRS) for $D_{\rm 2h}$ coronene, $D_{\rm 2h}$ buckminsterfullerene, and $C_{\rm 2v}$ Cobalt complex. We ran the calculations using 32 OpenMP threads, while between parenthesis we report the achieved FRS obtained with a serial calculation.}
    \label{tab:FRS}
\end{table}
We considered only the most expensive contractions, namely the ones that exibith a scaling of $O^3V^3$ and $O^2V^4$ (PPL). We note that the achieved FRS are sistematically larger for the $O^3V^3$ terms with respect to the PPL contraction. This can be explained since, even for the most efficient PPL algorithm (\textit{i.e.}, Algorithm~\ref{loop_a}), we need to perform two nested loops over the irreducible representations, which means that enforcing point group symmetry results in a larger number of significantly smaller matrix-matrix multiplications, with a loss of efficiency with respect to the $C_1$ treatment. On the other hand, the $O^3V^3$ terms are implemented in a way that only requires one explicit loop over the irreducible representations, affecting thus less the overall efficiency of their evaluation. This effect is less apparent for calculations that are run sequentially (see the numbers in parentheses in Table~\ref{tab:FRS}). 
Furthermore, we note that for the $C_{\rm 2v}$ cobalt complex, where only 4 irreducible representations are present, the achieved FRS are closer to the theoretical FRS. 

To show the full potential of our new implementation, we ran a calculation on a challenging system, namely buckminsterfullerene (C$_{60}$) with the \textit{cc}-pVTZ basis set,\cite{Dunning1989} which consists of 1800 basis functions (1620 virtual and 180 doubly occupied orbitals). The geometry can be downloaded from the Zenodo repository.\cite{nottoli_2023_ccsd_data} The computational complexity for this system is twofold since both the number of virtual orbitals and the number of electrons are high. We performed the calculation with the frozen-core approximation -- thus considering 120 occupied orbitals. We used a Cholesky decomposition threshold of 10$^{-4}$ and obtained 10169 CVs, which are nearly evenly distributed among the eight irreducible representations. The full set of CVs occupies only 29 GB of memory, which can be compared with the immense -- and perhaps prohibitive -- amount of memory, or rather disk space, that a traditional implementation would demand, which is about 3.4 TB for the full set of integrals $\bra{ab}\ket{cd}$ as estimated with a dry run. We note here that, in our implementation, to achieve maximum efficiency, we store the $\bra{ab}\ket{cd}$ integrals with $a\leq b$ and no other restriction. If one wanted to store only the unique elements of the $\bra{ab}\ket{cd}$ array, one would still need slightly less than 1TB of memory, which is still a formidable amount of memory that comes at the price of having to expand lower-triangular matrices into square matrices before performing the relevant matrix-matrix multiplications. 
To accelerate the convergence of the non-linear equation solver we exploited as usual the direct inversion in the iterative subspace (DIIS) strategy, and in order to avoid the in-core storage of the past expansion and error vectors we input/output them at each iteration. The memory required to store a set of amplitude ($t_{ij}^{ab}$) amounts in fact to approximately 30 GB, making it a potential bottleneck, especially when large DIIS expansion spaces are used. 

We first performed the computation using the ``loop ab'' PPL algorithm and distributing the work over 32 OpenMP threads. With this setup the total memory requested amounts to 478 GB, and the time for a single iteration is 6 hours 50 minutes proving the feasibility of the calculation. Furthermore, we note that using this algorithm the requested memory is almost independent of the number of threads used.  
We repeated the calculation using the ``loop a'' PPL algorithm and 12 OpenMP threads. Within this configuration however, the amount of memory requested is 1080 GB. A single iteration took about 2 hours, and the calculation converged in 1 day 19 hours to 10$^{-7}$ in the maximum norm of the residual. The PPL contraction is the most time consuming operation since about 65\% of a single-iteration time is spent in computing this term, while the rest is mostly spent in computing the $O^3V^3$ contributions. The converged CCSD energy is -2281.30020 E$_h$ of which -8.91574 E$_h$ represents the correlation contribution. 

In summary, we present a CCSD code that exploits Abelian point-group symmetry and the Cholesky decomposition of the two-electron integrals. The most expensive contribution -- namely the particle-particle ladder term -- has been implemented in two different versions using the symmetric/antisymmetric strategy. The two implementations differ in the memory required to store temporary arrays and the code is driven towards the best option depending on the available hardware resources. The implementation is suited to treat medium and large symmetric systems as demonstrated by performing CCSD calculations on buckminsterfullerene (C$_{60}$) with a polarized valence triple-zeta basis set. 

\begin{acknowledgments}
F.L. and T.N. acknowledge financial support from ICSC-Centro Nazionale di Ricerca in High Performance Computing, Big Data, and Quantum Computing, funded by the European Union -- Next Generation EU -- PNRR, Missione 4 Componente 2 Investimento 1.4. In Mainz, this work was supported by the Deutsche Forschungsgemeinschaft (DFG) via project B5 within the TRR 146 
"Multiscale Simulation Methods for Soft Matter Systems".
\end{acknowledgments}

\section*{Data Availability Statement}
The molecular geometries used in the paper are included as supplementary data in the Zenodo repository.\cite{nottoli_2023_ccsd_data}

\bibliography{biblio}

\end{document}